\documentstyle{article}
\begin{document}
\title{The gravitational wave rocket}
\author{W.B.Bonnor and M.S.Piper\\ School of Mathematical Sciences\\Queen 
Mary and Westfield College\\
London E1 4NS}
\maketitle
\setlength{\parindent}{0.5in}
\begin{abstract}
Einstein's equations admit solutions corresponding to photon rockets.
In these a massive particle recoils because of the anisotropic emission
of photons.  In this paper we ask whether rocket motion can be powered only
by the emission of gravitational waves.  We use the double series
approximation method and show that this is possible.  A loss of mass
and gain in momentum arise in the second approximation because of the
emission of quadrupole and octupole waves.
\end{abstract}
\section{Introduction}
Recently there have appeared a number of investigations into exact solutions 
of Einstein's equations corresponding to photon rockets
[1]-[7].  In these a massive particle recoils because of the anisotropic
emission of photons.  One can ask whether a similar motion is possible
powered by the emission of gravitational waves rather than photons.
An answer to this question is probably implicit in work many years ago by
Newman and collaborators [8]-[12] who studied the motion of a gravitating
system subject to no external forces but taking into account its internal
structure.  Another related work is that of Hogan and Robinson [13]
on the deviation of a particle from geodesic motion
under the influence of its own gravitational radiation.
However these workers were not concerned explicitly with
rocket motion, and it is not clear to us how their results could
be applied to the problem.

In this paper we study the gravitational wave rocket, using an approximation 
method.
We start with the photon rocket
metric
\begin{eqnarray}
ds^{2}=(1-f^{2}r^{2}\sin^{2}\theta-2fr\cos\theta-2Mr^{-1})du^{2}+2dudr
\nonumber\\
-r^{2}(d\theta^{2}+\sin^{2}\theta d\phi^{2})-2fr^{2}\sin\theta du d\theta
\end{eqnarray}
with
\[ 0<r,\; 0\leq\theta\leq\pi,\; 0\leq\phi\leq2\pi,\;-\infty<u<\infty.\]
The coordinates will be numbered
\[x^{1}=r, \;x^{2}=\theta,\;x^{3}=\phi,\;x^{4}=u.\]
$M$ and $f$ are arbitrary functions of the retarded time $u$, to be 
interpreted as the mass
of a particle at the origin and its acceleration along the negative direction
of the polar axis.  If $M=0$ the spacetime is flat and the coordinate system
has an origin with acceleration $f$.  Metric (1) satisfies Einstein's equations
\begin{equation}
R^{ik}-(1/2)Rg^{ik}=-8\pi T^{ik}=-8\pi \mu l^{i}l^{k}
\end{equation}
where $\mu$ is a scalar function and $l^{i}$ is a null vector.  The right-hand
side of (2) is the energy tensor for the photon fluid driving the rocket.
To drive the rocket by gravitational waves we must get rid of this energy
tensor, so that we solve the vacuum equations
\begin{equation}
R^{ik}=0
\end{equation}
Accordingly, we try to add
terms to (1) so that the right-hand side of (2) disappears.

The quantities on the right-hand side of (2) are
\begin{equation}
l^{i}=\delta^{i}_{1},\:\mu=r^{-2}(2\dot{M}-6Mf\cos\theta);
\end{equation}
where an overdot means differentiation with respect to $u$.  In (4)
the expressions $2\dot{M}$ and $-6Mf\cos\theta$ refer respectively to the
loss of mass and the change of momentum by the rocket because of the 
emission of photons.
We wish to add terms to the metric (1) which cancel these,
so eliminating the photons which power the photon rocket.
From the theory of gravitational radiation we have clues as to how to do this.
We know from many investigations that a source loses mass by the emission
of quadrupole waves [14], and that it gains momentum from recoil when it
emits quadrupole and octupole waves [15]-[19].  This suggests that the terms
we must add to (1) are those describing quadrupole and octupole gravitational 
waves.
If we do this correctly we should get equations of motion for the rocket in
terms of the gravitational radiation it emits: $M$ and $f$ will no longer
be arbitrary functions.

We shall add the quadrupole and octupole wave terms in the linear 
approximation,
but the cancellation of $T^{ik}$ in (2) will occur in the quadrupole-quadrupole
and quadrupole-octupole approximations.

In Section 2 we describe the metric and approximation method which we use, and
in Section 3 we explain the method of solving the field equations.  Sections
4-7 outline the solutions of various approximation steps, the main results
appearing in equations (31) and (34).  There is a brief concluding Section 8.

\section{The approximation method}
We assume axial symmetry and take the metric in the form
\begin{equation}
ds^{2}=-r^{2}(B d\theta^{2}+C\sin^{2}\theta d\phi^{2})+Ddu^{2}+2Fdrdu +2rG 
d\theta du;\;C=B^{-1},
\end{equation}
where $B,C,D,F,G$ are functions of $r,\theta,u.$  This metric, which is a 
form of that due to
Bondi [20], is particularly suitable for describing outgoing radiation.
It will be noticed that (1) is of the form (5).

We use the double-series approximation method which has been used previously
in conjunction with the metric (5) to study
the emission of gravitational waves from isolated sources [19], so we assume
that the solution we are looking for can be expanded in powers of constant 
parameters $m$, the
initial mass of the rocket, and $a$, some convenient length associated with 
it.

We envisage some internal vibrations of the rocket which generate varying
quadrupole and octupole moments $Q(u)$ and $O(u)$ about the symmetry axis, 
given by
\begin{equation}
Q=ma^{2}h(u),\: O=ma^{3}k(u)
\end{equation}
where $h$ and $k$ are smooth functions which vanish except for a finite
interval $0\leq u\leq u_{1}$.  There may, of course, be
multipole moments higher than the octupole, but we shall not be concerned
with them here.

The expansion of the metric functions takes the form
\begin{equation}
g_{ik}=\sum_{p=0}^{\infty}\sum_{s=0}^{\infty}m^{p}a^{s}\stackrel{(ps)}{g_{ik}},
\end{equation}
and
we assume that $M$ and $f$ in (1) can also be expanded in powers of $m$ and 
$a$:
\begin{equation}
M=\sum_{p=1}^{\infty}\sum_{a=0}^{\infty}m^{p}a^{s}\stackrel{(ps)}{M},\;\;
f=\sum_{p=1}^{\infty}\sum_{a=0}^{\infty}m^{p}a^{s}\stackrel{(ps)}{f}.
\end{equation}
In the above the $\stackrel{(ps)}{g_{ik}}, \stackrel{(ps)}{M}, 
\stackrel{(ps)}{f}$
are independent of $m$ and $a$,  and $\stackrel{(ps)}{M}, \stackrel{(ps)}{f}$
are functions of $u$ only.  In fact, since the starting value of $M$ is $m$
and decrements to the mass appear at higher approximations, we have
$\stackrel{(10)}{M}=1$.

Introducing the notation of (5) we write the metric coefficients as
\begin{eqnarray}
-g_{22}&=&r^{2}B=r^{2}[1+\sum_{p=1}^{\infty}\sum_{s=0}^{\infty}m^{p}a^{s}
\stackrel{(ps)}{B}],\\
-g_{33}&=&r^{2}\sin^{2}\theta C=r^{2}\sin^{2}\theta[1+\sum_{p=1}^{\infty}
\sum_{s=0}^{\infty}m^{p}a^{s}\stackrel{(ps)}{C}],\\
 g_{44}&=&D=1+\sum_{p=1}^{\infty}\sum_{s=0}^{\infty}m^{p}a^{s}
\stackrel{(ps)}{D},\\
 g_{14}&=&F=1+\sum_{p=1}^{\infty}\sum_{s=0}^{\infty}m^{p}a^{s}
\stackrel{(ps)}{F},\\
 g_{24}&=&rG=r\sum_{p=1}^{\infty}\sum_{s=0}^{\infty}m^{p}a^{s}
\stackrel{(ps)}{G}.
\end{eqnarray}
These summations ensure that if the mass parameter $m$ is put zero we have
flat spacetime in the form
\begin{equation}
ds^{2}=-r^{2}(d\theta^{2} + \sin^{2}\theta d\phi^{2}) +2drdu+du^{2}.
\end{equation}

We emphasise that we are looking for a metric of form (1) which satisfies the
{\em vacuum} equations (3).  There is now no energy tensor such as occurred
in the photon rocket.

If we substitute (7) into the field equations (3) we can separate out the
coefficients of $m^{p}a^{s}$.  Equating these to zero, we obtain a doubly
infinite set of second order differential equations.  The set which is the 
coefficient
of $m^{p}a^{s}$ will be called the $(ps)$ approximation.
The (00) approximation vanishes.  The (1s)
approximations are linear and homogeneous in the $\stackrel{(1s)}{g}$, and it
is here that we shall insert the quadrupole and octupole wave terms which
are of order $(12)$ and $(13)$ respectively.  The non-linear terms are those 
with
$p>1$.  For further description of the double-series method see [19].

For the reason described in Section 1, we expect changes of mass to appear
at the quadrupole-quadrupole, i.e. the (12)$\times$(12) or (24) approximation.
Changes in momentum should come from the quadrupole-octupole, i.e. 
(12)$\times$(13)
or (25) approximation; the acceleration engendered by this change will be
of order (15).  We accordingly take the initial terms in (8) to be
\begin{eqnarray}
M=m+m^{2}a^{4}\stackrel{(24)}{M},\\
f=m a^{5}\stackrel{(15)}{f}.
\end{eqnarray}

\section{Solution of the field equations}
The $(ps)$ approximation described above consists of
seven equations of the form
\begin{equation}
\Phi_{lm}(\stackrel{(ps)}{g_{ik}})=\Psi_{lm}(\stackrel{(qr)}{g_{ik}}),
\end{equation}
where the left hand side is linear in the $\stackrel{(ps)}{g_{ik}}$ (and their
derivatives), and the right hand side is non-linear in the
$\stackrel{(qr)}{g_{ik}}, (q<p, r\leq s)$ and is known from previous
approximation steps.  For instance, in the (24) approximation, which we shall 
consider in
Section 6, the terms on the right hand side could come from the following
combinations
\begin{eqnarray}
\stackrel{(10)}{g_{ik}}\times\stackrel{(14)}{g_{ik}}, 
\stackrel{(11)}{g_{ik}}\times
\stackrel{(13)}{g_{ik}}, \stackrel{(12)}{g_{ik}}\times
\stackrel{(12)}{g_{ik}}
\end{eqnarray}
(and their derivatives).

In Appendix I of [19] were given the explicit forms of the left hand sides 
of (17),
together with an algorithm for its solution.  Here we give only a summary
of the results.  Where there is no ambiguity we shall drop the superscripts
$(ps)$ to save printing.

The solution of the $(ps)$ approximation depends on an inhomogeneous pseudo
wave equation for $D$:
\begin{equation}
\Box^{\prime}D\equiv D_{11}-2D_{14}+2r^{-1}(D_{1}+D_{4})+r^{-2}(D_{22}+D_{2} 
\cot\theta)
=\Pi +2r^{-2}\chi_{4}(u,\theta)
\end{equation}
where subscripts 1,2,4 mean differentiation with respect to $r, \theta, u$,
where $\Pi$ consists of non-linear terms of order $(ps)$ known from previous
approximations and also terms in $\stackrel{(ps)}{F}$ which is known by a
simple integration of one of the $(ps)$ equations.  
$\stackrel{(ps)}{\chi(u,\theta)}$
is an arbitrary function of integration, and such a function arises at each
approximation step; it enters not only the equation for $D$, but also, directly
or indirectly, the equations for $B$ and $G$.
It plays an essential part in the construction of physically
sensible solutions.

The algorithm in [19] shows how to calculate $\stackrel{(ps)}{B}, 
\stackrel{(ps)}{C},
\stackrel{(ps)}{G}$ once $\stackrel{(ps)}{D}$ has been got from (19).  In
obtaining them one must ensure that there is no physical singularity on the
rotation axis $\theta=0, \theta=\pi$.  Sufficient conditions for this are that
\begin{equation}
B/\sin^{2}\theta,\;C/\sin^{2}\theta,\;D,\;F,\;G/\sin\theta
\end{equation}
shall be of class $C^{2}$ near $\sin\theta=0$ [20].
\section{The $(1s)$ approximations}
The $(10)$ approximation is simply the Schwarzschild solution in Bondi
coordinates.  In the notation we have described it is (omitting the $(10)$
superscript)
\begin{equation}
D=-2r^{-1},\;\;B=C=F=G=0.
\end{equation}

It was shown in [19] that the $(11)$ approximation allows no non-singular
solution except motion with constant velocity, which we shall put zero,
so abolishing the $(11)$ terms.  This corresponds with the known result
that there are known linear dipole gravitational waves.  However, a non-linear
dipole wave does appear at the $(25)$ approximation, as we shall see in
Section 7.

The $(12)$ and $(13)$ approximations, corresponding to quadrupole and
octupole waves were given in [19]:
\begin{eqnarray}
\stackrel{(12)}{B}=-\stackrel{(12)}{C}&=&\frac{1}{2}\sin^{2}\theta(r^{-1}
\ddot{h}+r^{-3}h),\\
\stackrel{(12)}{D}&=&-P_{2}(2r^{-1}\ddot{h}+2r^{-2}\dot{h}+r^{-3}h),\\
\stackrel{(12)}{F}&=&0,\\
\stackrel{(12)}{G}&=&\frac{1}{6}P_{2}^{\prime}(2r^{-1}\ddot{h}-4r^{-2}\dot{h}
-3r^{-3}h).
\end{eqnarray}

\begin{eqnarray}
\stackrel{(13)}{B}=-\stackrel{(13)}{C}&=&\frac{1}{12}\sin^{2}\theta\cos
\theta(2r^{-1}\stackrel{...}{k}+10r^{-3}\dot{k}+15r^{-4}k,\\
\stackrel{(13)}{D}&=&-\frac{1}{3}P_{3}(2r^{-1}\stackrel{...}{k}+4r^{-2}
\ddot{k}+5r^{-3}\dot{k}+3r^{-4}k),\\
\stackrel{(13)}{F}&=&0,\\
\stackrel{(13)}{G}&=&\frac{1}{36}P_{3}^{\prime}(2r^{-1}\stackrel{...}{k}
-8r^{-2}\ddot{k}-15r^{-3}\dot{k}-12r^{-4}k).
\end{eqnarray}
Here $P_{2}(\cos\theta), P_{3}(\cos\theta)$ are Legendre polynomials, 
$\prime$ means $d/d\theta$ and
$h,k$ are the functions introduced in (6).

The only other $(1s)$, i.e. linear, approximation which enters the work is
the $(15)$ one, which comes in because of $\stackrel{(15)}{f}(u)$.
We see from (1) and (16) that the $(15)$ terms in the metric are (dropping the
superscripts)
\begin{equation}
B=C=F=0,\;D=-2fr\cos\theta,\;G=-fr\sin\theta.
\end{equation}
If we substitute these expressions into (17) (with right-hand side zero
because we are considering a linear approximation) we find it satisfied
identically.\footnote {The terms (30) will give rise to a $(2\:10)$
approximation, as expected because of the appearance of $f^{2}$ in (1).
This too vanishes identically.  The satisfaction of the $(15)$
and $(2\:10)$ approximations occurs because (1) with $M=0$ is flat.}

\section{The second approximations}
Our methodology in this work, like that in [19] and [21], is to introduce the 
multipole
waves in the linear approximation only.  That means we do not add any further 
solutions
of the homogeneous equations (i.e. (19) with $\Psi_{lm}=0$) in the higher
approximations.  This paradigm is ambiguous, because the solution
of the non-linear equations (i.e. (19) with $\Psi_{lm}\neq0$) is not unique.
This non-uniqueness of the double series method has been discussed in [21].
It applies to other approximation methods which are governed by a set of
equations such as (19), since at any approximation stage one can always add a
solution of the homogeneous equations.  Fortunately the non-uniqueness does 
not trouble us
in the work described in this paper.

Turning to the individual $(2s)$ steps, we find there are no non-linear terms
of order $(20)$ or $(21)$, so we choose $\stackrel{(20)}{g_{ik}}$ and
$\stackrel{(21)}{g_{ik}}$ both zero.  There are terms of order $(22)$ and 
$(23)$
but, as shown in [19], they refer to wave tails, which die away gradually to 
zero,
and have no effect on the motion of the rocket.

The first step of interest here is the $(24)$ one, to which we now turn.

\section{The $(24)$ approximation}
The possible non-linear contributions in the general $(24)$ approximation are 
those in equation (18).
However, in our work we are taking $\stackrel{(11)}{g_{ik}}$ equal to zero, 
and
inserting no $(14)$ multipole, so the only non-linear terms we need are those
corresponding to $\stackrel{(12)}{g_{ik}}\times\stackrel{(12)}{g_{ik}}$.  
This case was
completely solved in [21]; the solution is lengthy and will not be reproduced
here.  It is free from singularities on the rotation axis,
and consists of oscillating terms which cease when the internal motions cease,
together with certain terms showing secular change.  The latter can be removed
by coordinate transformations, as was shown in [19], except for one
term in $\stackrel{(24)}{D}$ which refers to a change in the
Schwarzschild mass of the rocket, corresponding to the transmission of energy
by the gravitational waves.  This is to be identified with $\stackrel{(24)}{M}$
in equation (15):
\begin{equation}
\stackrel{(24)}{M}=-\frac{1}{30}\int_{-\infty}^{u}\stackrel{...}{h}^{2}du.
\end{equation}
{\em Eqn (31) gives the mass-loss of the rocket in terms of its quadrupole
oscillations.}  It
agrees with emission of energy according to the quadrupole formula.

\section{The $(25)$ approximation}
In solving the $(25)$ approximation two conditions will be imposed.  First,
there must be no singularities on the rotation axis.  Sufficient conditions
for this were stated in equation (20).  In the solution we are considering 
these can be satisfied by suitable choice
of arbitrary functions, such as $\chi(u,\theta)$, arising in the algorithm
for solving each $(ps)$ approximation.  In general the question of whether
$\chi$ can be chosen to avoid singularities along the axis of symmetry
depends on the satisfaction of a particular relationship between the non-
linear terms at each order of the approximation.  It has been shown that this
relationship can be satisfied up to the fourth order in $m$ and $a$ [22].

Secondly, we must ensure that the rocket remains at the origin $r=0$
so that we know that its acceleration is $f$.  This will be so if its dipole 
moment about the
origin vanishes at all time.  We shall assume that the dipole moment enters
the $(25)$ approximation as the coefficient of $r^{-2}\cos\theta$ in the
expansion of $\stackrel{(25)}{D}$, as it would into the expansion of the
Newtonian gravitational potential.  It will be noted that the generating
equation (19), in the linear approximation and the static case, (in which
$\Pi$ and $\chi_{4}$ are zero) reduces to Laplace's equation with axial
symmetry, and has a solution $\mu r^{-2}\cos\theta$ for a dipole $\mu$.

The non-linear terms in the $(25)$ approximation come from the following
combinations
\[\stackrel{(10)}{g_{ik}}\times\stackrel{(15)}{g_{ik}},\;\;
\stackrel{(12)}{g_{ik}}\times\stackrel{(13)}{g_{ik}}.\]
The former of these consists of a single expression
\[-6m^{2}a^{5}\stackrel{(15)}{f} \cos\theta/r^{2} \]
arising in the Ricci tensor component $R^{11}$; in fact it corresponds to
part of (4) which we wish to cancel, as explained in Section 1.  The second 
combination
arises from the quadrupole-octupole interaction.

The $(25)$ approximation has been completely solved by one of us (MSP).
Unfortunately the solution is extremely long (the expression for
$\stackrel{(25)}{D}$ alone contains 66 terms) so we do not print it here,
but we will gladly send a copy to anybody who asks for it.  The  following
is a brief summary.

$\stackrel{(25)}{D}$ can be expressed exactly in a finite series
\[\stackrel{(25)}{D}=\sum_{n=1}^{7}r^{-n}\stackrel{n}{\Delta}(u,\theta).\]
The arbitrary functions which arise in solving the $(25)$ step, including 
$\stackrel{(25)}{\chi}(u,\theta)$ which enters the
pseudo wave equation (19), must be chosen so that $\stackrel{(25)}{G}$
and $\stackrel{(25)}{B}$ are regular on the axis.  When this is done, we find
\begin{eqnarray}
\stackrel{(25)}{D}=\cos\theta[r^{-1}(\frac{2}{105}Y-6\int_{0}^{u}(
\stackrel{(15)}{f(x)}dx)\nonumber\\
+r^{-2}(-2\int_{0}^{u}\int_{0}^{x}(\stackrel{(15)}{f(y)}dydx+\frac{2}{315}Z 
+\frac{2}{105}E-\frac{1}{105}p(u)q(u))
+A(r,u)]\nonumber\\
+P_{3} X(r,u)  +P_{5} W(r,u).
\end{eqnarray}
In this
\[p(x):=d^{2}h(x)/dx^{2},\;q(x):=d^{3}k(x)/dx^{3},\]
\begin{equation}
Y=\int _{0}^{u}\frac{dp(x)}{dx}\frac{dq(x)}{dx}dx,\;Z=\int_{0}^{u}Y(x)dx,\;
E=\int_{0}^{u}\frac{dp(x)}{dx}q(x)dx,
\end{equation}
$A$ denotes regular transient terms of order $r^{-3}$ and higher, and $W,X$
are finite series in $r^{-1}$ with regular transient coefficients which
need not concern us here.

We now impose the second condition referred to above, namely that the 
coefficient
of $r^{-2}\cos\theta$ shall vanish.  Differentiating this twice with respect 
to $u$ we have
\[\stackrel{(15)}{f}=\frac{1}{630}(2\dot{p}\dot{q}-3p\ddot{q}+3q\ddot{p}),\]
where $p,q$ are functions of $u$.  So using equation (16) and introducing 
units of
customary dimensions
\begin{equation}
f=\frac{mGa^{5}}{630c^{7}}(2\dot{p}\dot{q}-3p\ddot{q}+3q\ddot{p}),
\end{equation}
$G$ being here the gravitational constant.
{\em This gives the acceleration of the rocket in terms of the rates of
change of the quadrupole and octupole moments induced by the rocket's
internal motions} correct to order $(15)$ in the expansion parameters $m$ 
and $a$.

The term in $r^{-1}\cos\theta$ in $\stackrel{(25)}{D}$ is
\[\frac{\cos\theta}{35r}(p\dot{q}-q\dot{p}),\]
which represents a dipole gravitational wave.  Thus although a dipole
wave is forbidden in the linear approximation, one arises at the $(25)$ stage.

The solutions for $\stackrel{(25)}{B}$ and $\stackrel{(25)}{G}$ obtained by
the algorithm in [19] contain secular terms involving
integrals like those occurring in equation (32).  However, they can be 
removed by a
coordinate transformation, namely the $\alpha$-transformation given in
[20], which was also used in [19], eqn (10.14).

The acceleration lasts while there are internal motions, i.e. from $u=0$
till $u=u_{1}$.  Integrating the acceleration (34) between these limits
we have the velocity $V$ acquired, assuming the rocket was at rest at $u=0$:
\[ V=\frac{mGa^{5}}{315c^{7}}\int_{0}^{u_{1}}\dot{p}\dot{q}du,\]
agreeing with a result obtained in [19].

\section{Conclusion}
The main results of the paper are equations (31) and (34), giving 
respectively the
mass-loss and acceleration of the rocket in terms of its internal motions 
represented
in our work by variations of the quadrupole and octupole moments.

We have considered the following approximations
\[(00), (10), (12), (13), (15), (24), (25), (2\:10)\]
and referred briefly to the wave-tail approximations $(22)$ and $(23)$
which do not affect the equations of mass-loss and motion of the rocket.
There will of course be other approximation steps, even if we confine
ourselves to the quadrupole and octupole oscillations.  For example,
the $(26)$ approximation arising from $(13)\times(13)$ terms will
contribute to the \vspace{0.2in} mass-loss.\\

{\sc\large REFERENCES}\\
{[1]}Bonnor W B 1994 {\em Class. Quantum Grav.} {\bf11} 2007\\
{[2]}Damour T 1995 {\em Class. Quantum Grav.} {\bf12} 725\\
{[3]}Bonnor W B 1996 {\em Class. Quantum Grav.} {\bf13} 277\\
{[4]}Dain S, Moreschi O M and Gleiser R J 1996 {\em Class. Quantum Grav.} 
{\bf13} 1155\\
{[5[}Cornish F H J and Micklewright B 1996 {\em Class. Quantum Grav.} {\bf13} 
2499\\
{[6]}Cornish F H J and Micklewright B 1996 {\em Class. Quantum Grav.} {\bf13} 
2505\\
{[7]}Hellaby C 1996 {\em Class. Quantum Grav.} {\bf13} 2537\\
{[8]}Newman E T and Posadas R 1969 {\em Phys. Rev. Lett.} {\bf22} 1196\\
{[9]}Newman E T and Posadas R 1969 {\em Phys. Rev.} {\bf187} 1784\\
{[10]}Newman E T and Posadas R 1971 {\em J. Math. Phys.} {\bf12} 2319\\
{[11]}Lind R W, Messmer J and Newman E T 1972 {\em Phys. Rev. Lett.} {\bf28} 
857\\
{[12]}Lind R W , Messmer Jand Newman E T 1972 {\em J. Math. Phys.} {\bf13} 
1884\\
{[13]}Hogan P A and Robinson I 1986 {\em Foundations of Physics} {\bf16} 
455\\
{[14]}Misner C W, Thorne K S and Wheeler J 1970 {\em Gravitation} (Freeman) 
page 989\\
{[15]}Bonnor W B and Rotenberg M A 1961 {\em Proc. R. Soc.} A{\bf265} 109\\
{[16]}Papapetrou A 1962 {\em C.R. Acad. Sci. Paris} {\bf255} 1578\\
{[17]}Peres A 1962 {\em Phys. Rev.} {\bf128} 2471\\
{[18]}Cooperstock F I {\em Ap. J.} {\bf213} 250\\
{[19]}Bonnor W B and Rotenberg M A 1966 {\em Proc. R. Soc.} A{\bf289} 247\\
{[20]}Bondi H, van der Burg M G J and Metzner A W K 1962 {\em Proc. R. Soc.} 
A{\bf269} 21\\
{[21]}Hunter A J and Rotenberg M A 1969 {\em J. Phys. A (Gen. Phys.)} {\bf2} 
34\\
{[22]}Piper M S 1997 {\em Class. Quantum Grav.} (in press)

\end{document}